\documentclass[12pt]{article}
\usepackage{epsfig}
\title{ \bf Spectral Density of (Pseudo)Scalar Currents at Finite Temperature}

\author{El\c{s}en Veli Veliev $^{*1}$, Kazem Azizi $^{\dag2}$, Hayriye Sundu $^{*3}$, G\"{u}l\c{s}ah
Kaya $^{*4}$
\\ $^{*}$ Physics Department, Kocaeli University, Umuttepe Yerle\c{s}kesi \\
41380 Izmit, Turkey \\
 $^{\dag}$ Physics Division, Faculty of Arts and
Sciences, Do\v{g}u\c{s} University, \\ Ac{\i}badem-Kad{\i}k\"{o}y
\\
 $^1$ e-mail:elsen@kocaeli.edu.tr\\
$^2$e-mail:kazizi@dogus.edu.tr\\
$^3$email:hayriye.sundu@kocaeli.edu.tr\\
$^4$email:gulsahbozkir@kocaeli.edu.tr}
\date{}
\begin{document}
\setlength{\baselineskip}{24pt}
\maketitle
\setlength{\baselineskip}{7mm}
\begin{abstract}
We study the spectral densities of (pseudo)scalar currents at finite
temperature in general case when mass of two quarks are different.
Such spectral densities are necessary for the phenomenological
investigation of hadronic parameters. We use quark propagator at
finite temperature and show that an additional branch cut arises in
spectral density, which corresponds to particle absorption from the
medium. The obtained results at $T\rightarrow0$ limit are in good
agreement with the vacuum results.

\end{abstract}

\setcounter{page}{1}
\section{Introduction}
Heavy ion collision experiments provide an opportunity to
investigate particle properties in the medium. Inspired by these
experiments, there is an increasing interest to investigate the
properties of hadronic matter under extreme condition
\cite{1},\cite{2}. In general, the media created by these collisions
consist  different mesons and baryons. Investigation of hadronic
properties at finite temperature and density directly using the
fundamental thermal QCD Lagrangian is highly desirable. However,
such interactions occur in a region very far from the perturbative
regime, where the quark-gluon coupling constant becomes large and
the perturbative methods is not suitable for calculation of these
properties. Therefore, we need nonperturbative approaches.

Some nonperturbative approaches are Lattice QCD, Heavy Quark
Effective Theory (HQET), QCD Sum Rules, etc. Among these approaches,
the QCD Sum Rules method  \cite{3} and its extension to finite
temperature  \cite{4} has been widely used as an efficient and
applicable tool to investigate the  hadronic properties  \cite{5}.
The main  elements in QCD sum rules approach are correlation
function and dispersion relation. The correlation function is an
object with dual nature. At large negative  $q^2$, it can be
evaluated by perturbative methods, whereas at positive $q^2$, it
must be represented in terms of hadronic observables. Dispersion
relation allows us to link the thermal correlator for positive
values of $q^2$ to the one for negative values. Spectral functions
in different cases were studied in the literature
\cite{6}-\cite{13}.

In the present work, we investigate the two-point thermal correlator
using the real time formulation of the thermal field theory
\cite{14}. We calculate the spectral densities of (pseudo)scalar
currents at finite temperature, which are necessary for the
phenomenological investigation of (pseudo)scalar mesons parameters.
We use quark propagator at finite temperature and show how an
additional branch cut corresponding to particle absorption from the
medium arises in spectral density. We also compare our results with
predictions obtained at zero temperature.

\section{Thermal Spectral Densities of (Pseudo) Scalar Currents }
We begin by considering the thermal correlation function,
\begin{equation}\label{eqn1}
\Pi(q, T)=i \int d^{4}x  e^{iq\cdot x} \Big{\langle} {\cal T} \Big{(}J(x)J^{+}(0)\Big{)}\Big{\rangle}, \\
\end{equation}
where $J(x)=:\bar{q}_{1}(x)\Gamma q_{2}(x):$ is the interpolating
current that carries the quantum numbers of the state concerned and
${\cal T}$ indicates  the time ordered product. Here $\Gamma=I$ or
$i\gamma_5$ for scalar and pseudoscalar particles, respectively. The
thermal average of the operator, $A={\cal T}(J(x)J^{+}(0))$,
appearing in the above thermal correlator is expressed as
\begin{equation}\label{eqn2}
\langle A\rangle=Tr e^{-\beta H}A/Tr e^{-\beta H}, \\
\end{equation}
where $H$ is the QCD Hamiltonian, and $\beta=1/T$ stands for the
inverse of the temperature. Traces are carried out over any complete
set of states. In the real time version, thermal correlator has the
form of a $2\times2$ matrix. However, this matrix may be
diagonalized, when it is expressed by a single analytic function,
which determines completely the dynamics of the corresponding
two-point function \cite{15}. As this function is simply related to
any one, say the $11$-component of the matrix, we need to calculate
only this component of the correlation function. The $11$-component
of the thermal quark propagator is a sum of vacuum quark propagator
expression and a term depending on the Fermi distribution function,
\begin{equation}\label{eqn3}
S(q)=(\gamma^{\mu}q_{\mu}+m)\Big{(}\frac{1}{q^{2}-m^{2}+i\varepsilon}+2\pi
i n(|q_0|)\delta(q^{2}-m^{2})\Big{)}, \\
\end{equation}
where $n(x)$ is the Fermi distribution function, $n(x)=[\exp(\beta
x)+1]^{-1}$. Now, we proceed to obtain the temperature-dependent
dispersion relation. The time ordering product in Eq. (1) can be
expressed as
\begin{equation}\label{eqn4}
\langle T(J(x)J^{+}(x'))\rangle=\theta(x_{0}-x'_{0})\langle J(x) J^{+}(x')\rangle+\theta(x'_{0}-x_{0})\langle J^{+}(x')J(x)\rangle, \\
\end{equation}
where $\theta(x)$ is step function.

Using Kubo-Martin-Schwinger relation, $\langle J(x_{0})
J^{+}(x'_{0})\rangle=\langle J^{+}(x'_{0})J(x_{0}+i\beta)\rangle$
for thermal expectation and making Fourier and some other
transformations, we get the following expression for the thermal
correlation function in momentum space \cite{14}:
\begin{equation}\label{eqn5}
\Pi (|\textbf{q}|, q_{0})=\frac{1}{2\pi}\int_{-\infty}^{\infty}dq'_{0}M(|\textbf{q}|, q'_{0})\Big{(}\frac{1}{q_{0}-q'_{0}+i\varepsilon}-\frac{\exp(-\beta q_{0})}{q_{0}-q'_{0}-i\varepsilon}\Big{)} , \\
\end{equation}
where
\begin{equation}\label{eqn6}
M(|\textbf{q}|, q_{0})=\int d^{4}x e^{i q\cdot x}\langle J(x) J^{+}(0)\rangle . \\
\end{equation}
In the above transformations, the following standard integral
representation for the  $\theta$- step function  is used:
\begin{equation}\label{eqn7}
\theta(x_{0}-x'_{0})=\frac{1}{2 i \pi}\int_{-\infty}^{\infty}dk_{0}\frac{\exp[ik_{0}(x_{0}-x'_{0})]}{k_{0}-i\varepsilon} . \\
\end{equation}
The imaginary part of the correlation function can be simply
evaluated using the formula $\frac{i}{x+i \varepsilon}=\pi
\delta(x)+i P (\frac{1}{x})$, which leads to \cite{16}:
\begin{equation}\label{eqn8}
\Pi (q,T)=\int_{0}^{\infty}ds \frac{\rho (s)}{s+Q_{0}^{2}} , \\
\end{equation}
where $\rho (q,T)=\frac{1}{\pi} Im \Pi(q,T) \tanh \frac{\beta
q_{0}}{2}$ and $Q_{0}^{2}=-q_{0}^{2}$. In some cases, the
correlation function has ultraviolet divergent. If the spectral
density does not vanish at $s\rightarrow\infty$, the dispersion
integral in Eq. (8) diverges. A standard way to overcome this
problem is to subtract first few terms of its Taylor expansion at
$q^{2}=0$ from $\Pi(q,T)$. The thermal correlation function in
momentum space can be written as
\begin{equation}\label{eqn9}
\Pi(q,T)=-i \int \frac{d^{4}k}{(2\pi)^{4}}Tr (\Gamma S(k) \Gamma S(k-q)) , \\
\end{equation}
where $\Gamma=I$ and $i\gamma_{5}$  for scalar and pseudoscalar
particles, respectively. Inserting propagators from Eq. (3) in Eq.
(9) and carrying out the $k_{0}$ integration, we obtain the
imaginary part of $\Pi(q,T)$ in the following form:
\begin{eqnarray}\label{eqn10}
&&Im\Pi(q,T)=-N_{c}\int\frac{d\textbf{k}}{8\pi^{2}}\frac{1}{\omega_{1}\omega_{2}}\Big{[}(\omega_{1}^{2}-\textbf{k}^{2}+\textbf{k}\cdot
\textbf{q}-\omega_{1}q_{0}\pm m_{1}m_{2}) \nonumber\\&& \times
[(1-n_{1}-n_{2}+2n_{1}n_{2})
\delta(q_{0}-\omega_{1}-\omega_{2})-(n_{1}+n_{2}-2n_{1}n_{2})\delta(q_{0}-\omega_{1}+\omega_{2})]
\nonumber\\&&+(\omega_{1}^{2}-\textbf{k}^{2}+\textbf{k}\cdot
\textbf{q}+\omega_{1}q_{0}\pm m_{1}m_{2}) \nonumber\\&& \times
[(1-n_{1}-n_{2}+2n_{1}n_{2})
\delta(q_{0}+\omega_{1}+\omega_{2})-(n_{1}+n_{2}-2n_{1}n_{2})\delta(q_{0}+\omega_{1}-\omega_{2})]\Big{]}
,
\end{eqnarray}
where $m_{1}$ and $m_{2}$ are quark masses,
$\omega_{1}=\sqrt{\textbf{k}^{2}+m_{1}^{2}}$ ,
$\omega_{2}=\sqrt{(\textbf{k-q})^{2}+m_{2}^{2}}$ ,
$n_{1}=n(\omega_{1})$, $n_{2}=n(\omega_{2})$ and the plus and minus
signs in front of $m_{1}$, $m_{2}$, correspond to scalar and
pseudoscalar particles, respectively. The term, which does not
include the Fermi distribution functions, show the vacuum
contribution. Terms including the Fermi distributions depict medium
contributions. The delta-functions in the different terms of Eq.
(10) control the regions of non-vanishing imaginary parts of
$\Pi(q,T)$ , which define the position of the branch cuts. As seen
the term including $\delta(q_{0}-\omega_{1}-\omega_{2})$ gives
contribution when $q_{0}=\omega_{1}+\omega_{2}$ . Using
Cauchy-Schwarz inequality, $( \sum_{i=1}^{{n}}
a_{i}^{2})(\sum_{i=1}^{n}b_{i}^{2})\geq(\sum_{i=1}^{n}a_{i}
b_{i})^{2}$ we see that,
\begin{equation}\label{eqn11}
\omega_{1}\omega_{2}=\sqrt{\textbf{k}^{2}+m_{1}^{2}}\sqrt{(\textbf{k-q})^{2}+m_{2}^{2}}\geq|\textbf{k}| |\textbf{k-q}|+m_{1} m_{2} ,\\
\end{equation}
and for $q_{0}=\omega_{1}+\omega_{2}$, we get,
\begin{equation}\label{eqn12}
q_{0}^{2}=m_{1}^{2}+\textbf{k}^{2}+m_{2}^{2}+(\textbf{k-q})^{2}+2\omega_{1}\omega_{2}\geq (m_{1}+m_{2})^{2}+\textbf{q}^{2} .\\
\end{equation}
Therefore, we obtain the first branch cut, $q^{2}\geq
(m_{1}+m_{2})^{2}$, which coincides with zero temperature cut
describing the standard threshold for particle decays. This term
survives at zero temperature and it is called the annihilation term.
On the other hand, the term including
$\delta(q_{0}-\omega_{1}+\omega_{2})$ gives contribution when
$q_{0}=\omega_{1}-\omega_{2}$. Similarly to the above expression, we
obtain,
\begin{equation}\label{eqn13}
q_{0}^{2}=m_{1}^{2}+\textbf{k}^{2}+m_{2}^{2}+(\textbf{k-q})^{2}-2\omega_{1}\omega_{2}\leq (m_{1}-m_{2})^{2}+\textbf{q}^{2}  ,\\
\end{equation}
and therefore an additional branch cut arises  at finite
temperature, $q^{2}\leq (m_{1}-m_{2})^{2} $, which corresponds  to
particle absorption from the medium. It is called scattering term
and vanishes  at $T=0$.

In the following, we restrict our calculations with
$|\textbf{q}|=0$, when there is no angular dependence. Note that,
with $|\textbf{q}|=0$, the value of $|\textbf{k}|$, fixed by the
$\delta$-functions in Eq. (10), is the magnitude of three momentum
of quark or antiquark in the center-of-mass of the quark-antiquark
system:
\begin{equation}\label{eqn14}
\textbf{k}^{2}=\frac{\Big{(}q_{0}^{2}-(m_{1}+m_{2})^{2}\Big{)}\Big{(}q_{0}^{2}-(m_{1}-m_{2})^{2}\Big{)}}{4 q_{0}^{2}} .\\
\end{equation}
In the $|\textbf{q}|=0$ case, as it is seen, the terms including
$\delta(q_{0}-\omega_{1}-\omega_{2})$ and
$\delta(q_{0}+\omega_{1}+\omega_{2})$ functions in Eq. (10), give
contributions at the regions, $q_{0}\geq (m_{1}+m_{2})$ and
$q_{0}\leq -(m_{1}+m_{2})$, respectively giving the vacuum cuts.
Similarly, the terms including $\delta(q_{0}-\omega_{1}+\omega_{2})$
and $\delta(q_{0}+\omega_{1}-\omega_{2})$  functions  in Eq. (10),
give contributions at in the regions, $0\leq q_{0}\leq
(m_{1}-m_{2})$ and $-(m_{1}-m_{2})\leq q_{0}\leq 0 $, respectively
giving the Landau cuts. After straightforward calculations, we find
the vacuum part of the $Im \Pi (q,T)$ as:
\begin{equation}\label{eqn15}
Im \Pi (q_{0},T=0)=\frac{N_{c}}{8 \pi q_{0}^{2}}\sqrt{(q_{0}^{2}-m_{1}^{2}-m_{2}^{2})^{2}-4m_{1}^{2} m_{2}^{2}}\Big{(}q_{0}^{2}-(m_{1}-m_{2})^{2}\Big{)} .\\
\end{equation}
Taking into account both branch cuts after some transformations, the
annihilation and scattering parts of spectral density is found as:
\begin{equation}\label{eqn16}
\rho_{a,pert}(s,T)=\rho_{0}(s)\Big{[}1-n\Big{(}\frac{\sqrt{s}}{2}\Big{(}1+\frac{m_{1}^{2}-m_{2}^{2}}{s}\Big{)}\Big{)}-n\Big{(}\frac{\sqrt{s}}{2}\Big{(}1-
\frac{m_{1}^{2}-m_{2}^{2}}{s}\Big{)}\Big{)}\Big{]} ,\\
\end{equation}
for $(m_{1}+m_{2})^{2}\leq s\leq\ \infty $,
\begin{equation}\label{eqn17}
\rho_{s,pert}(s,T)=\rho_{0}(s)\Big{[}n\Big{(}\frac{\sqrt{s}}{2}\Big{(}1+\frac{m_{1}^{2}-m_{2}^{2}}{s}\Big{)}\Big{)}-n\Big{(}-\frac{\sqrt{s}}{2}\Big{(}1-
\frac{m_{1}^{2}-m_{2}^{2}}{s}\Big{)}\Big{)}\Big{]} ,\\
\end{equation}
for $0\leq s \leq (m_{1}-m_{2})^{2}$, with $m_{1}\geq m_{2}$. Here,
$\rho _{0}(s)$ is the spectral density in the lowest order of
perturbation theory at zero temperature and it is given by
\begin{equation}\label{eqn18}
\rho_{0}(s)=\frac{3}{8\pi^{2}s}q^{2}(s)v^{n}(s),\\
\end{equation}
where $q(s)=s-(m_{1}-m_{2})^{2}$ and $v(s)=\Big{(}1-4m_{1}
m_{2}/q(s)\Big{)}^{1/2}$. Here $n=3$ and $n=1$ for scalar and
pseudoscalar particles, respectively. As it is seen, at
$T\rightarrow0$ limit these expressions are in good consistency with
the vacuum expressions. Moreover, the obtained results  are well
consistent with the existing results in $m_{1}=m_{2}$ case
\cite{17}-\cite{19} for the scalar and pseudoscalar particles.

As an example, we present the dependence of the annihilation and
scattering parts of the spectral density for $K^{\pm}$ and $D^{\pm}$
particles  in Figs. 1 and 2. In numerical analysis, we use the
values $m_{s}=0,13$ GeV and $m_{c}=1,46 $ GeV for the quark masses.
As it is clear, in the region of the standard threshold  for
particle decays, the $\rho_{0}(s)$ is replaced by the annihilation
term. In the case of light mesons,  the values of
$\rho_{a,pert}(s,T)$ considerably differ from those of the
$\rho_{0}(s)$. However, in the case of heavy mesons, the
$\rho_{a,pert}(s,T)$ and $\rho_{0}(s)$ values are very close to each
other. From Fig. 1,  we also see that the in light $K^{\pm}$ cases,
the medium contributions play important role and consist higher
percentage of the total value.

Our concluding result is that  the thermal contributions contribute
significantly to the spectral function.

%

\section{Acknowledgement}
The authors would like to thank T. M. Aliev for his useful
discussions. This work are supported in part by the Scientific and
Technological Research Council of Turkey (TUBITAK) under the
research project No. 110T284 and in part by Kocaeli University under
the research project No. 2010/32.
\begin{figure}[h]
\centerline{\epsfig{figure=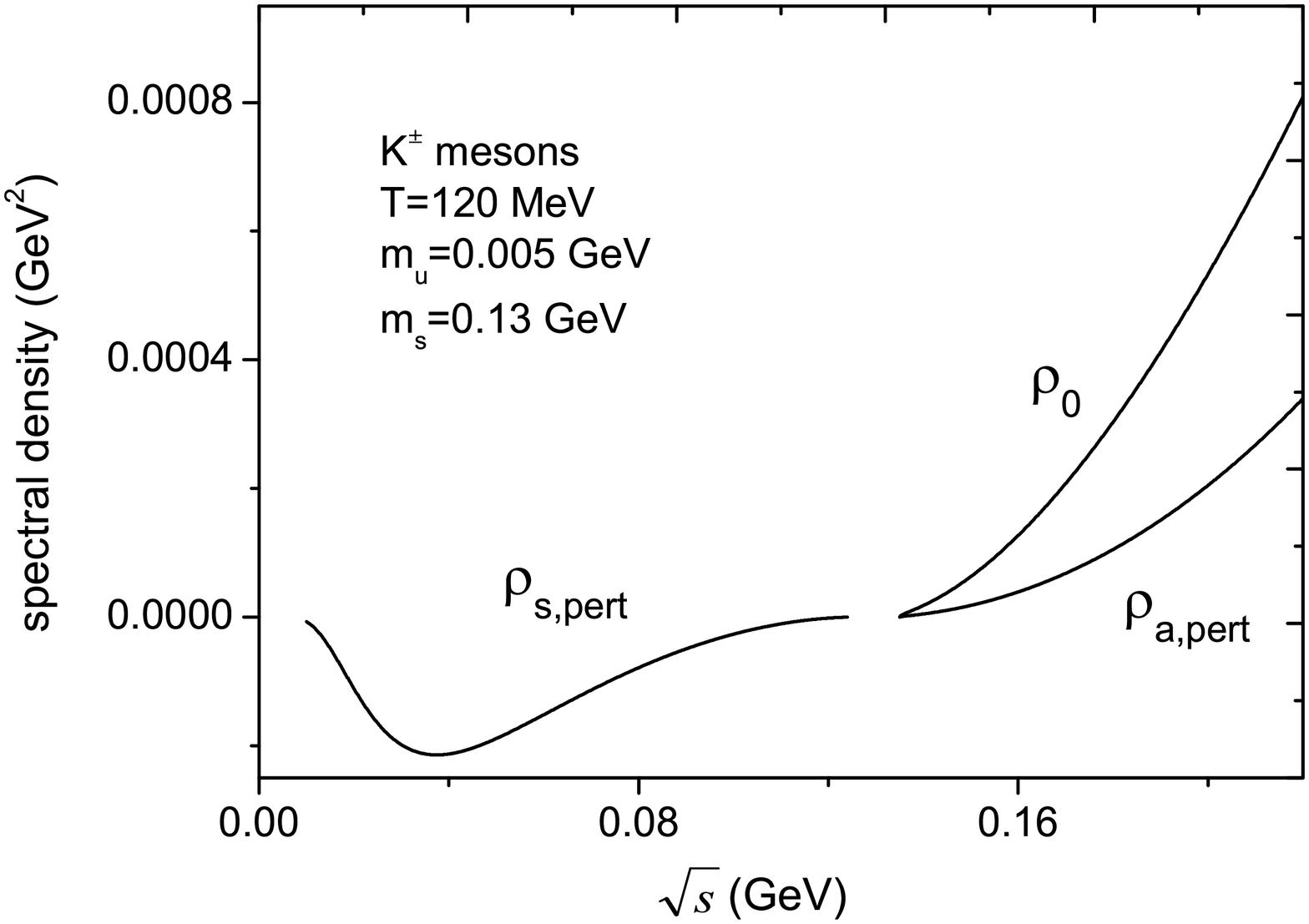,height=80mm}}\caption{ The
dependence of the spectral density of $K^{\pm}$ meson at temperature
$T=120$ MeV on the $\sqrt{s}$ parameter. } \label{BRmh0}
\end{figure}
\begin{figure}[h]
\centerline{\epsfig{figure=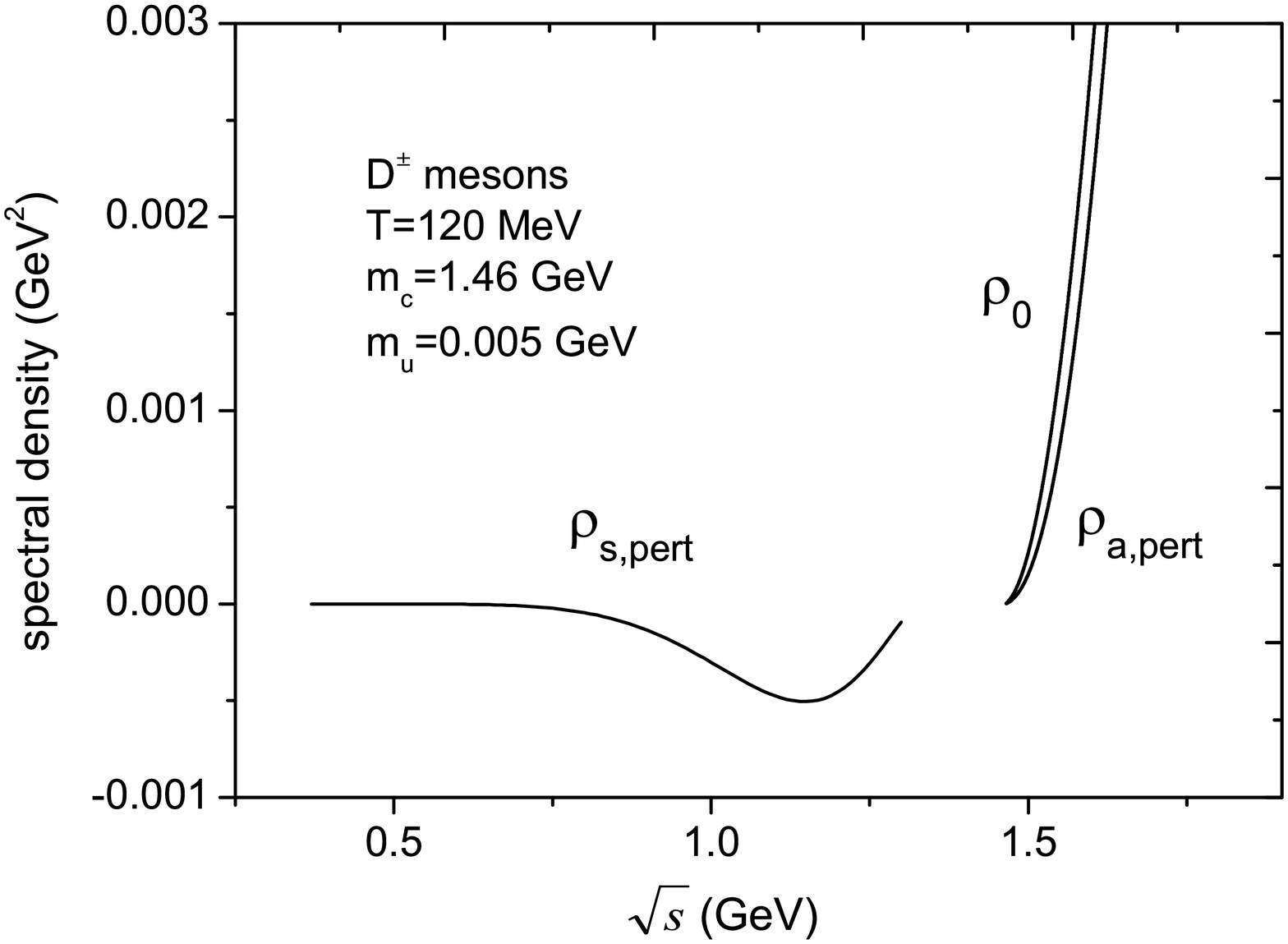,height=80mm}}\caption{ The
dependence of the spectral density of $D^{\pm}$ meson at temperature
$T=120$ MeV on the $\sqrt{s}$ parameter.} \label{BRmh0}
\end{figure}
\end{document}